\documentclass{article}

\usepackage{arxiv}

\usepackage{amssymb}
\usepackage{amsmath}
\usepackage{algorithm}
\usepackage{algpseudocode}

\usepackage{algorithm}
\usepackage{algpseudocode}

\usepackage[hyphens]{url}
\usepackage{float}
\usepackage{mathtools}
\usepackage{threeparttable}
\usepackage{booktabs}

\title{Efficient Nonlinear Multiscale Prediction for Unseen Polycrystalline Textures via Self-Supervised Microstructure Pretraining}

\author{
 Ting-Ju Wei \\
  Department of Civil Engineering\\
  National Taiwan University\\
  Taipei, Taiwan \\
   \And
 Chuin-Shan Chen\thanks{Corresponding author. Email: \texttt{dchen@ntu.edu.tw}} \\
  Department of Civil Engineering\\
  Department of Materials Science and Engineering\\
  National Taiwan University\\
  Taipei, Taiwan \\
}

\begin{document}
\maketitle
\begin{abstract}
Predicting the nonlinear mechanical response of polycrystalline materials across diverse crystallographic textures remains computationally prohibitive, and existing reduced-order surrogates are typically fit to a single microstructural realization, precluding reuse on unseen textures. We address both limitations through a self-supervised pretraining strategy. A three-dimensional masked autoencoder is pretrained on 100,000 voxelized synthetic face-centered cubic (FCC) microstructures whose textures systematically span the texture hull via hierarchical simplex sampling, yielding transferable, texture-aware latent representations. A differentiable homogenization operator then maps these representations to the parameters of an orientation-aware interaction-based deep material network (ODMN). Given a previously unseen microstructure, the pretrained encoder infers a standalone ODMN that, coupled with crystal plasticity, reproduces the nonlinear loading–unloading–reloading stress–strain response with mean relative error below 2\%, at a 304$\times$ CPU-time speedup over full-field crystal-plasticity direct numerical simulation. The pretrained representation is also highly data-efficient: on a label-limited homogenized-stiffness regression task, pretraining raises the validation R² from below 0.1 (trained from scratch) to above 0.8. Together, these results demonstrate that self-supervised pretraining yields physically meaningful, transferable microstructural representations and provides a scalable framework for microstructure–property inference. The present scope (FCC systems with equiaxed grains) is a deliberate first step, with extensions to morphological texture and other crystal systems outlined.
\end{abstract}


\section{Introduction}\label{sec1}

Polycrystalline materials are central to many engineering applications, and their macroscopic mechanical behavior is governed by the underlying microstructural architecture, in particular the crystallographic texture~\cite{motaman2020anisotropic}. Establishing microstructure--property relationships is therefore a central objective in materials design and performance optimization. Advances in artificial intelligence (AI) have opened new opportunities for building data-driven links between microstructural descriptors and effective material responses, offering a route toward accelerated computational materials modeling~\cite{su2022multiscale}.

A range of machine learning architectures has been explored for polycrystalline systems. Convolutional neural networks (CNNs), such as U-Net, have been used to predict full-field stress distributions in viscoplastic polycrystals~\cite{khorrami2023artificial}. Graph neural networks (GNNs) capture grain-level topological interactions and have been used to predict the effective magnetostriction of heterogeneous polycrystals~\cite{dai2021graph}. Variational autoencoders (VAEs) learn compact latent representations of electron backscatter diffraction (EBSD) patterns, yielding embeddings that improve EBSD indexing efficiency~\cite{liu2025learning}. Transformer-based generative adversarial networks have been used to synthesize three-dimensional bioinspired microstructures~\cite{chiang2023generating}. More recently, masked autoencoders (MAEs) have been applied to synthetic two-dimensional polycrystalline datasets for classification; although pretraining was restricted to randomly textured microstructures, these studies indicate the potential of self-supervised learning for texture-aware representation learning~\cite{belamri2025quaternion}. These efforts point to the value of latent representation learning for microstructure-aware materials modeling~\cite{su2023modelfree}.

The concept of a foundation model extends this line of work: large-scale pretraining on abundant data yields latent representations that transfer across diverse downstream tasks. In natural language processing, large language models such as BERT and GPT exemplify this paradigm by learning highly generalizable embeddings from massive text corpora~\cite{devlin2018bert,radford2019language}. Analogously, foundation models for materials science have begun to emerge. For example, a masked autoencoder pretrained on synthetic two-dimensional composite microstructures has been applied across several downstream tasks~\cite{WEI2025114397}, and large pretrained vision models have been fine-tuned for materials image analysis~\cite{cheng2026pore}.

To efficiently predict nonlinear mechanical responses of polycrystalline materials, physically informed surrogate models have been developed as alternatives to direct full-field simulations. Among these approaches, the orientation-aware interaction-based deep material network (ODMN) has been proposed as an efficient and physically interpretable surrogate modeling framework~\cite{Wei01}. The ODMN extends the deep material network (DMN) architecture~\cite{liu2019deep,noels2022micromechanics,noels2022interaction,wei2026dmn} and incorporates crystallographic texture effects through a hierarchical reduced-order representation. A distinctive feature of the ODMN is that its parameters can be trained using linear elastic homogenized responses, while remaining capable of extrapolating to nonlinear mechanical behavior during online prediction~\cite{liu2019deep,gajek2020micromechanics,noels2022micromechanics,wan2024decoding}. However, ODMN parameters are tied to individual microstructural realizations, limiting transferability across different textures. To address this issue, several studies have proposed interpolation- or learning-based strategies to generalize DMN parameters across microstructures~\cite{huang2022microstructure,jean2024graph,wu2026convolutional, WEI2025114397}. Nevertheless, extension of such approaches to three-dimensional polycrystalline microstructures, particularly for mapping learned representations to ODMN parameters, remains largely unexplored.

Two considerations clarify what a learned representation adds beyond existing descriptors and solvers. First, although modern homogenization solvers based on the fast Fourier transform (FFT) can solve a single $45^3$ elastic representative volume element (RVE) in seconds, the cost that the present framework amortizes is the repeated nonlinear crystal-plasticity evaluation of many previously unseen textures: once an ODMN is inferred for a microstructure, its online evaluation is far cheaper than full-field crystal-plasticity direct numerical simulation, which we quantify through a cost--benefit analysis (Section~\ref{sec:cost}). Second, relative to prior masked-autoencoder studies on two-dimensional or randomly textured microstructures~\cite{belamri2025quaternion, WEI2025114397} and to deep material networks whose parameters are tied to a single microstructural realization~\cite{liu2019deep, Wei01}, the present contribution is the coupling of large-scale three-dimensional self-supervised pretraining with a differentiable homogenization operator, enabling \emph{transferable} inference of physically interpretable surrogate parameters for previously unseen polycrystals.

Motivated by these developments, this study investigates foundation models as a unifying framework for crystallographic texture informatics in polycrystalline materials. We use the term \emph{foundation model} in the specific sense of large-scale self-supervised pretraining that yields transferable representations; establishing generality across crystal systems and experimental data is beyond the present scope and is examined in Section~\ref{sec:limits}. We introduce a foundation model as a universal microstructure encoder whose learned latent representations are systematically mapped to ODMN parameters, enabling transferable prediction of nonlinear responses for previously unseen microstructures. The model is pretrained on a large-scale synthetic dataset spanning the complete crystallographic texture space of face-centered cubic (FCC) crystals, generated via hierarchical simplex sampling (HSS)~\cite{johnson2018efficient,nino2024evolution}. Voxel-based polycrystalline microstructures constructed from these textures are used for large-scale self-supervised pretraining.

Using this dataset, we develop a three-dimensional (3D), voxel-based foundation model built on a masked autoencoder architecture that learns texture-aware latent representations in a self-supervised manner. The pretrained model is subsequently fine-tuned for two representative downstream tasks: (1) prediction of homogenized stiffness and (2) inference of material surrogate model parameters via linear projection of the learned latent representations. In the second task, the inferred surrogate parameters are evaluated under crystal plasticity to predict nonlinear homogenized responses. Across both tasks, the pretrained model consistently outperforms baselines trained from scratch, demonstrating improved generalization and confirming the value of large-scale self-supervised pretraining for texture-aware representation learning.

The remainder of this paper is organized as follows. Section~\ref{sec2} details the proposed methodology, including dataset generation, foundation model pretraining, and downstream fine-tuning strategies. Section~\ref{sec3} presents and discusses the numerical results for homogenized stiffness prediction and nonlinear response modeling. Finally, Section~\ref{sec4} summarizes the key findings and outlines perspectives for future developments in foundation model-based approaches for polycrystalline materials.

\section{Methods}\label{sec2}

\subsection{The pretraining dataset}\label{sec21}

The pretraining dataset is designed to systematically explore the crystallographic texture space of FCC crystals. To this end, it is first necessary to define the sampling space of polycrystalline microstructures in terms of their admissible crystallographic textures. When a texture is represented by an orientation distribution function (ODF), it can be approximated using a discrete basis of Dirac delta functions~\cite{fullwood2010microstructure,fast2008application,johnson2016texture,johnson2018efficient}. In this formulation, the ODF is expressed as
\begin{equation}\label{eq:odf_definition}
f(q) \approx \sum_{j=1}^{J} p_j \, \delta(q, q^{(j)}),
\end{equation}
where $q \in \mathrm{SO}(3)$ denotes the crystallographic orientation, with $\mathrm{SO}(3)$ the special orthogonal group of three-dimensional rotations; $\delta(\cdot,\cdot)$ is the Dirac delta function defined on the orientation manifold; $q^{(j)}$ represents a discrete set of fundamental orientations obtained from a discretization of $\mathrm{SO}(3)$; and $p_j$ denotes the probability weight associated with each orientation state, with $J$ the total number of discrete orientations. In this study, the fundamental orientations are uniformly sampled from the FCC fundamental zone with an angular resolution of $10^{\circ}$, resulting in $J = 618$ discrete orientations in Eq.~\eqref{eq:odf_definition}.

The texture coefficients $\{p_j\}$ can be interpreted as probabilistic weights that quantify the likelihood of occurrence of each fundamental orientation. Accordingly, the ODF can be represented in vector form as $\vec{p} = (p_1, p_2, \ldots, p_J)$. The convex set spanned by all admissible vectors $\vec{p}$ defines the texture hull $M_H$~\cite{johnson2018efficient},
\begin{equation}
M_H = \left\{ \vec{p} \,\middle|\, 0 \le p_j \le 1,\; \sum_{j=1}^{J} p_j = 1 \right\}.
\end{equation}

\noindent Each point within the texture hull corresponds to a realizable crystallographic texture. Owing to the high dimensionality of this convex space, an efficient sampling strategy is required to ensure adequate coverage. In this work, HSS is employed to systematically generate representative ODFs by sampling points $\vec{p}$ within the texture hull~\cite{johnson2018efficient,nino2024evolution}. A total of $100{,}000$ ODF instances are generated, each representing a distinct crystallographic texture used for constructing the pretraining dataset.

Following ODF generation, each sampled texture is realized as an RVE using DREAM3D-NX~\cite{groeber2014dream}. Each RVE consists of a $45 \times 45 \times 45$ voxel grid containing approximately 810 equiaxed grains. The voxel-wise crystallographic orientations are subsequently reduced to the fundamental zone and converted to quaternion representations to ensure rotational continuity during self-supervised pretraining. The chosen $45\times45\times45$ resolution with approximately 810 grains balances microstructural representativeness against the computational cost of large-scale pretraining, consistent with established guidance on representative-volume-element size for polycrystalline elasticity~\cite{savvas2016determination}.

\subsection{Self-supervised pretraining stage}\label{sec:pretraining}

\begin{figure}[ht]
    \centering
    \includegraphics[width=\linewidth]{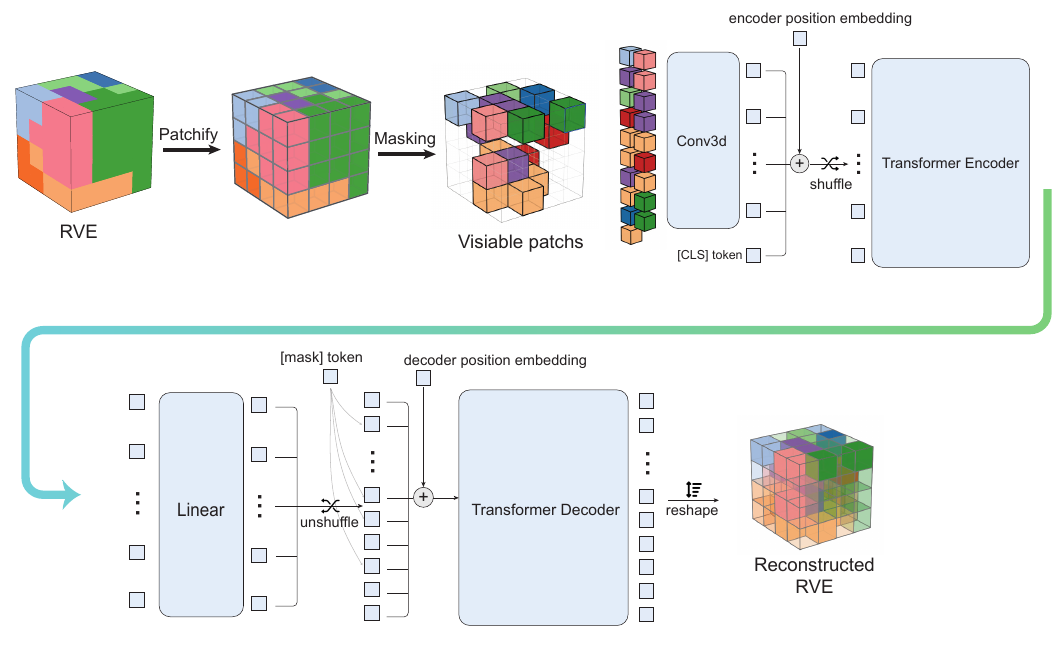}
    \caption{Schematic illustration of the proposed polycrystal foundation model during the self-supervised pretraining stage. The colored voxel grids within the RVE symbolically represent different crystallographic orientations and are shown for illustrative purposes only.}
    \label{fig:mmae_architecture_pretrain}
    
\end{figure}

During pretraining, each voxel-based RVE is represented as a tensor of size $45 \times 45 \times 45 \times 4$, where the final dimension corresponds to the quaternion components encoding local crystallographic orientations within the fundamental zone. The input tensor is partitioned into non-overlapping cubic patches of size $\mathcal{P} \times \mathcal{P} \times \mathcal{P}$ voxels with $\mathcal{P} = 9$, yielding $N_p = 125$ patches per training instance. A random subset of patches is masked at a prescribed masking ratio (e.g., 40\%), resulting in 75 visible and 50 masked patches.

Each visible patch is mapped into a latent embedding space via a 3D convolutional projection layer with kernel size and stride $(9, 9, 9)$, producing 768-dimensional feature vectors. These patch embeddings are then processed by a transformer-based encoder consisting of $\mathcal{N}_e = 12$ blocks, each with an embedding dimension $\mathcal{D}_e = 768$ and $\mathcal{H}_e = 12$ self-attention heads. The encoder outputs compact latent representations that capture texture-aware statistical features of the polycrystalline microstructure.

A lightweight transformer decoder is subsequently employed to reconstruct the masked patches. The decoder comprises $\mathcal{N}_d = 8$ blocks with embedding dimension $\mathcal{D}_d = 512$ and $\mathcal{H}_d = 16$ attention heads. Its role is to infer the quaternion fields of the masked patches from contextual information provided by the visible ones, thereby enforcing spatial and orientational coherence in the reconstructed microstructure.

To encode spatial information, three-dimensional sinusoidal positional encodings (PEs) are added to the inputs of both the encoder and decoder following the formulations in~\cite{vaswani2017attention, WEI2025114397}. After patchification, the patch grid has dimensions $\mathcal{G} \times \mathcal{G} \times \mathcal{G}$, with $\mathcal{G} = 5$ in this study. Each spatial coordinate $(x,y,z)$ is independently encoded using sinusoidal functions at multiple frequencies.
The embedding dimension $\mathcal{D}_e$ is evenly partitioned across the three spatial
axes, such that
$\mathcal{D}_x = \mathcal{D}_y = \mathcal{D}_z = \mathcal{D}_e / 3$.
The axis-wise positional encodings are defined as

\begin{equation}
\begin{aligned}
\text{PE}_x^{(2i)}(x) &= \sin\left( \frac{x}{10000^{2i / \mathcal{D}_x}} \right), \quad
\text{PE}_x^{(2i+1)}(x) = \cos\left( \frac{x}{10000^{2i / \mathcal{D}_x}} \right), \\
\text{PE}_y^{(2i)}(y) &= \sin\left( \frac{y}{10000^{2i / \mathcal{D}_y}} \right), \quad
\text{PE}_y^{(2i+1)}(y) = \cos\left( \frac{y}{10000^{2i / \mathcal{D}_y}} \right), \\
\text{PE}_z^{(2i)}(z) &= \sin\left( \frac{z}{10000^{2i / \mathcal{D}_z}} \right), \quad
\text{PE}_z^{(2i+1)}(z) = \cos\left( \frac{z}{10000^{2i / \mathcal{D}_z}} \right),
\end{aligned}
\end{equation}

\noindent where $i = 0, 1, \ldots, \mathcal{D}_x/2 - 1$.

In addition, a learnable classification token, denoted as [CLS], is prepended to the embedded patch sequence prior to transformer encoding. This token aggregates global microstructural information through the self-attention mechanism. Its positional encoding is set to zero, $\text{PE}_{\text{[CLS]}} = \mathbf{0}$. The complete positional encoding applied to the encoder input is therefore given by

\begin{equation}
\begin{aligned}
\mathbf{PE}_{\text{encoder}}
&= \text{stack}\Big(
\text{PE}_{\text{[CLS]}}, \\
&\quad
\left\{
\text{concat}\big(
\text{PE}_x(x),\, \text{PE}_y(y),\, \text{PE}_z(z)
\big)
\;\middle|\;
(x,y,z) \in \{0,\ldots,\mathcal{G}-1\}^3
\right\}
\Big) \\
&\in \mathbb{R}^{(1 + N_p) \times \mathcal{D}_e}.
\end{aligned}
\end{equation}

This asymmetric encoder--decoder architecture mitigates shortcut learning and compels the network to infer high-level statistical dependencies between visible and masked regions of the microstructure. The pretraining objective minimizes the mean squared error (MSE) between the reconstructed quaternion fields $\hat{\mathbf{Q}}_{\text{masked}}^{(i)}$ and the corresponding ground-truth fields $\mathbf{Q}_{\text{masked}}^{(i)}$ within the masked patches. Each patch has a feature dimensionality $d = \mathcal{P} \times \mathcal{P} \times \mathcal{P} \times C$, where $C$ denotes the number of quaternion components. The pretraining loss is defined as

\begin{equation}
    \mathcal{L}_{\text{pretrain}} = \frac{1}{|\mathcal{M}|} \sum_{i \in \mathcal{M}} \frac{1}{d} \left\| \hat{\mathbf{Q}}_{\text{masked}}^{(i)} - \mathbf{Q}_{\text{masked}}^{(i)} \right\|_2^2,
\end{equation}
where $\mathcal{M}$ denotes the set of masked patch indices. This objective enforces accurate reconstruction of masked quaternion fields and guides the model to learn microstructural representations essential for downstream tasks.

\subsection{Training configuration}\label{sec:training_config}

Both pretraining and downstream fine-tuning are performed using mini-batch stochastic gradient-based optimization. During pretraining, the encoder is trained with the masked-reconstruction objective over the full corpus of $100{,}000$ microstructures. For each downstream task, the encoder is initialized from the pretrained weights and jointly optimized with a lightweight task-specific head by minimizing the corresponding supervised loss. The checkpoint with the lowest validation loss is retained for evaluation.

The optimization settings used in each stage are summarized in Table~\ref{tab:training_config}. The base learning rate $\eta_{\mathrm{base}}$ is defined with respect to a reference batch size of 256. For a batch size $B$, the peak learning rate $\eta_{\mathrm{peak}}$ follows the linear scaling rule,
\begin{equation}
    \eta_{\mathrm{peak}}
    =
    \eta_{\mathrm{base}}\times \frac{B}{256}.
    \label{eq:linear_lr_scaling}
\end{equation}

The learning rate is scheduled using a linear warmup followed by a half-cycle cosine decay. Let $t$ denote the current epoch, $t_w$ the number of warmup epochs, $T$ the total number of epochs, $\eta_{\mathrm{peak}}$ the peak learning rate after warmup, and $\eta_{\min}$ the minimum learning rate. The learning rate at epoch $t$ is given by
\begin{equation}
\eta(t)=
\begin{cases}
\displaystyle
\eta_{\mathrm{peak}}\frac{t}{t_w},
& 0 \leq t < t_w, \\[8pt]
\displaystyle
\eta_{\min}
+
\frac{1}{2}
\left(
\eta_{\mathrm{peak}}-\eta_{\min}
\right)
\left[
1+\cos\left(
\pi\frac{t-t_w}{T-t_w}
\right)
\right],
& t_w \leq t \leq T .
\end{cases}
\label{eq:warmup_cosine_schedule}
\end{equation}

\begin{table}[htbp]
\centering
\caption{Training configuration for pretraining and downstream fine-tuning.}
\label{tab:training_config}
\begin{tabular}{@{}llll@{}}
\toprule
Setting & Pretraining & Task~I & Task~II \\
\midrule
Optimizer          & AdamW & AdamW & AdamW \\
Base learning rate & $1.0\times10^{-4}$ & $1.0\times10^{-3}$ & $1.0\times10^{-4}$ \\
Learning-rate schedule & \multicolumn{3}{c}{Linear warmup followed by half-cycle cosine decay} \\
Weight decay       & $0.05$ & $0$ & $0$ \\
Batch size         & $768$ & $128$ & $768$ \\
Epochs             & $1400$ & $200$ & $70$ \\
Warmup             & $40$ epochs & $5$ epochs & $5$ epochs \\
Hardware           & NVIDIA H100 & NVIDIA RTX4090 & NVIDIA H100 \\
\bottomrule
\end{tabular}
\end{table}

\subsection{Downstream training}

Following self-supervised pretraining, the encoder is transferred to downstream tasks and fine-tuned end-to-end for task-specific objectives. In this stage, the pretrained encoder is used as a microstructure feature extractor, while lightweight task-specific heads are introduced to adapt the learned latent representations to different prediction targets. The learnable [CLS] token is employed as the global latent descriptor of the microstructure and is connected to a linear prediction head for the downstream tasks considered in this study.

\subsubsection{Downstream task I: Homogenized stiffness prediction}

\begin{figure}[ht]
    \centering
    \includegraphics[width=\linewidth]{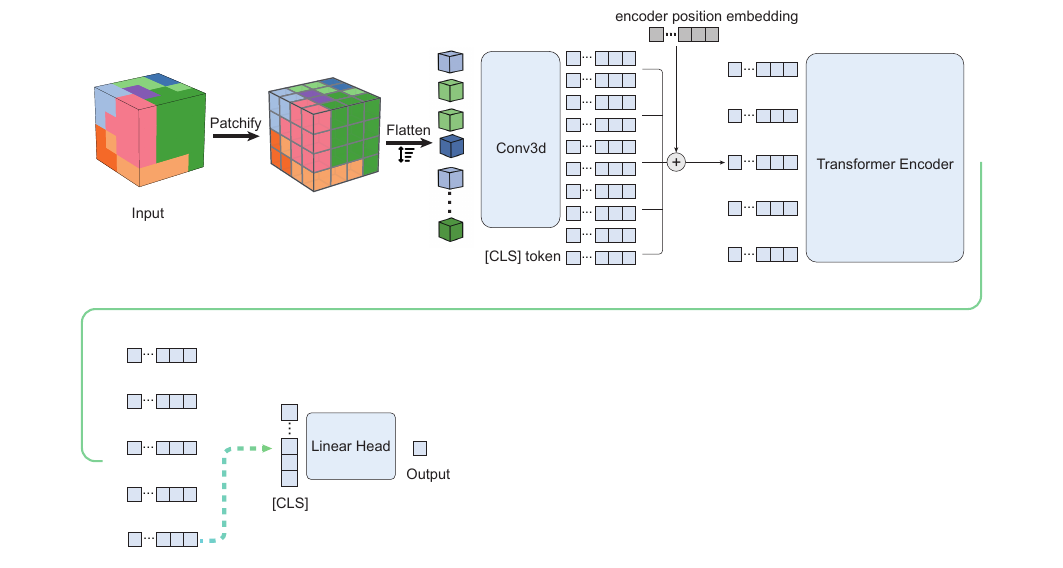}
    \caption{Schematic illustration of the downstream workflow for homogenized stiffness prediction. The pretrained encoder extracts texture-aware latent representations, which are subsequently mapped to homogenized stiffness components through a linear regression head.}
    
    \label{fig:stiffness_prediction}
\end{figure}

In polycrystalline materials, the homogenized mechanical response is predominantly governed by the underlying crystallographic texture, which dictates the collective anisotropy arising from the orientation distribution of constituent grains. This downstream task is designed to evaluate whether the latent representations learned during self-supervised pretraining can be effectively transferred to predict homogenized elastic properties.

Specifically, the model predicts three principal components of the homogenized stiffness tensor, namely $\bar{C}_{1111}$, $\bar{C}_{2222}$, and $\bar{C}_{3333}$, which correspond to the normal elastic moduli along the three orthogonal sample axes. Accurate prediction of these components provides a direct measure of the ability of the pretrained latent representations to map crystallographic texture information to effective macroscopic mechanical responses.

As illustrated in Fig.~\ref{fig:stiffness_prediction}, the pretrained 3D convolutional projection layer and transformer-based encoder are initialized from the pretraining stage and jointly fine-tuned during downstream training. The [CLS] token, which aggregates global microstructural information through the self-attention mechanism, serves as the latent representation of the RVE. A lightweight linear regression head is appended to this token to map the latent features to the target stiffness components.

For downstream evaluation, a total of 5,000 ODFs are sampled from the texture hull using the HSS algorithm described in Section~\ref{sec21}, ensuring adequate coverage of the crystallographic texture space. Each ODF is realized as a voxel-based RVE using DREAM3D-NX, consisting of approximately 810 equiaxed grains discretized on a $45 \times 45 \times 45$ voxel grid. The dataset is randomly split into 80\% for training and 20\% for validation to maintain balanced statistical coverage across the texture space. Homogenized stiffness tensors are computed using the DAMASK-FFT solver~\cite{roters2019damask} under periodic boundary conditions. Each grain is modeled as a single crystal with cubic elastic symmetry, characterized by the elastic stiffness constants listed in Table~\ref{tab:cubic_elasticity}.

\begin{table}[h]
    \centering
    \caption{Elastic stiffness constants of the cubic single crystal used in the downstream DAMASK-FFT simulations.}
    \label{tab:cubic_elasticity}
    \begin{tabular}{@{}ccc@{}}
        \toprule
        $C_{11}$ (GPa) & $C_{12}$ (GPa) & $C_{44}$ (GPa) \\ 
        \midrule
        107.3 & 60.8 & 28.3 \\
        \bottomrule
    \end{tabular}
\end{table}

\subsubsection{Formulation of the ODMN}

Following the motivation introduced in Section~\ref{sec1}, this section summarizes the ODMN formulation relevant to the present study. The ODMN is a physically interpretable surrogate model for polycrystalline materials that extends the DMN architecture~\cite{liu2019deep,noels2022micromechanics,noels2022interaction} with explicit incorporation of crystallographic texture effects~\cite{Wei01}.

Conceptually, the ODMN represents a polycrystalline microstructure as a parameterized reduced-order system that approximates the homogenized mechanical response of a full-field RVE~\cite{wu2025stochastic,shin2023deep}. Starting from a unit cell $\Omega$, the reduced-order microstructure is hierarchically partitioned into $2^N$ subdomains,
\begin{equation}
\Omega = \bigcup_{i=0}^{2^N-1} \Omega_i ,
\end{equation}
where each subdomain $\Omega_i$ corresponds to a material node $\mathcal{M}^{i}$ in the network. The network depth $N$ controls the resolution of the reduced-order representation, with larger $N$ enabling a more expressive approximation of the homogenized response.

The hierarchical partitioning is encoded by a binary-tree material network of depth $N$. At each internal tree node located at level $l$ and position $p$, an interaction direction $\vec{\mathbf{N}}^{\,l}_{p}$ is introduced to characterize the interface orientation governing stress equilibrium between adjacent subdomains. At each leaf-level material node $\mathcal{M}^{i}$, a set of parameters $\{ z^{i}, \alpha^{i}, \beta^{i}, \gamma^{i} \}$ is assigned to subdomain $\Omega_i$, where $z^{i}$ controls its volume fraction and $(\alpha^{i}, \beta^{i}, \gamma^{i})$ are the three Euler angles specifying its local crystallographic orientation.

The interaction direction $\vec{\mathbf{N}}$ is parameterized by two angular variables, $\theta$ and $\phi$, defining a unit normal vector according to
\begin{equation}\label{eq:direction_vector}
\vec{\mathbf{N}} =
\begin{bmatrix}
\cos(2\pi\phi)\sin(\pi\theta) \\
\sin(2\pi\phi)\sin(\pi\theta) \\
\cos(\pi\theta)
\end{bmatrix}.
\end{equation}

\noindent For an ODMN of depth $N$, the complete set of trainable parameters is given by
\begin{equation}\label{eq:trainable_parameters}
\begin{aligned}
\mathcal{F} =\;
& \left\{ z^i, \alpha^i, \beta^i, \gamma^i \;\middle|\; i = 0, 1, \dots, 2^N - 1 \right\} \\
& \cup \left\{ \theta^{l}_{p}, \phi^{l}_{p} \;\middle|\; l = 0, 1, \dots, N - 1,\; p = 0, 1, \dots, 2^l - 1 \right\},
\end{aligned}
\end{equation}
where the first set corresponds to material-node parameters and the second set specifies the interaction directions at each internal tree node.

A key property of the ODMN is that its parameters can be identified during an offline training stage using only linear elastic homogenized stiffness data, while the resulting network remains capable of extrapolating to nonlinear mechanical behavior during online prediction~\cite{liu2019deep,gajek2020micromechanics,noels2022micromechanics,wan2024decoding}. This capability arises because the learned parameters encode a first-order approximation of the microstructural mechanical response under linear elasticity, which is preserved within the hierarchical homogenization operators during online evaluation~\cite{gajek2020micromechanics}.

During online prediction, the ODMN performs a downscaling operation from the macroscopic deformation gradient $\bar{\mathbf{F}}$ to the deformation gradient $\mathbf{F}^i$ of each subdomain $\Omega_i$ according to
\begin{equation}
\mathbf{F}^i = \bar{\mathbf{F}} + \sum_{j=0}^{2^{N}-2} \alpha^{i,j} \, \mathbf{a}^j \otimes \mathbf{N}^j, \quad i = 0, 1, \dots, 2^{N}-1 ,
\label{eq:interaction_mapping}
\end{equation}
where the interaction coefficients $\alpha^{i,j}$ and interaction directions $\mathbf{N}^j$ are determined by the parameter set $\mathcal{F}$, and the interaction vectors $\mathbf{a}^j$ are fluctuation degrees of freedom introduced below.

At each material node, the local constitutive response is evaluated through
\begin{equation}
\mathbf{P}^i,\; \frac{\partial \mathbf{P}^i}{\partial \mathbf{F}^i} = \mathbb{P}^i\!\left( \mathbf{F}^i,\; \boldsymbol{\psi}^i \right),
\end{equation}
where $\mathbf{P}^i$ denotes the first Piola--Kirchhoff stress of the subdomain $\Omega_i$, $\partial \mathbf{P}^i / \partial \mathbf{F}^i$ is the corresponding consistent tangent operator, and $\mathbb{P}^i(\cdot)$ represents the local material law parameterized by the internal state variables $\boldsymbol{\psi}^i$ (e.g., history-dependent variables in crystal plasticity). The resulting local stresses and tangents are subsequently upscaled through the hierarchical network to recover the homogenized stress $\bar{\mathbf{P}}$ and its consistent tangent.

The interaction vectors $\mathbf{a}^j$ represent fluctuation fields associated with the interfaces between adjacent subdomains. During each Newton--Raphson iteration, the interaction vectors are initialized as $\mathbf{a}^j = \mathbf{0}$ and updated to satisfy the Hill--Mandel condition
\begin{equation}
\left( \sum_{i=0}^{2^N-1} W^i \, \mathbf{P}^i \, \alpha^{i,j} \right) \cdot \mathbf{N}^j = 0, \qquad j = 0, \ldots, 2^N - 2 ,
\end{equation}
where $W^i$ denotes the volume fraction of subdomain $\Omega_i$.

Similar to the original DMN framework, ODMN parameters are tied to specific microstructural realizations, which limits direct transferability across different microstructures. In the present work, this limitation is addressed by coupling the ODMN with a foundation model that provides transferable latent representations of polycrystalline microstructures. The resulting framework enables systematic inference of ODMN parameters for previously unseen three-dimensional microstructures, as described in the following subsection.

\subsubsection{Downstream task II: Nonlinear homogenized response prediction}

This downstream task addresses the limited transferability of ODMN parameters by coupling the ODMN with the pretrained foundation model. Specifically, the foundation model is employed to systematically infer ODMN parameters for previously unseen polycrystalline microstructures, thereby extending the pretrained encoder from linear-elastic property prediction to nonlinear homogenized response prediction~\cite{Wei01}. The inferred ODMN parameters are subsequently used to evaluate nonlinear stress--strain responses under crystal plasticity. This downstream task, therefore, assesses whether the learned latent representations can be transferred from elastic property prediction to the more challenging task of surrogate model parameter inference.

\begin{figure}[!ht]
\centering
\includegraphics[width=\linewidth]{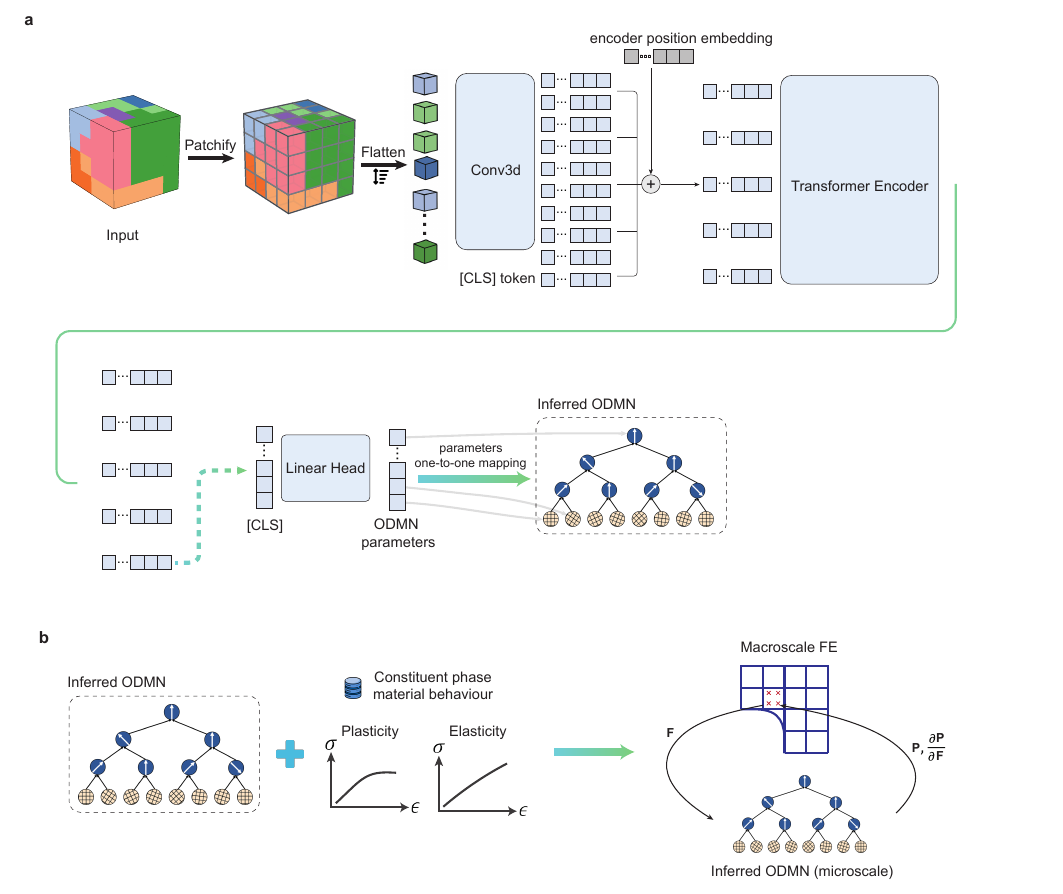}
\caption{Schematic illustration of the downstream workflow for nonlinear homogenized response prediction. 
(a) During the downstream offline training stage, the pretrained encoder extracts texture-aware latent features, which are subsequently passed through a linear regression head to predict the ODMN parameters. 
(b) During the online prediction stage, the inferred ODMN is coupled with the constituent phase material behavior to enable nonlinear homogenized response prediction.}
\label{fig:odmn_downstream}
\end{figure}

In this work, voxel-based RVEs are used as inputs to the pretrained 3D convolutional and transformer-based encoder. A linear regression head is appended to the [CLS] token to predict the parameters of the ODMN, as illustrated in Fig.~\ref{fig:odmn_downstream}(a). The ODMN employed as the surrogate model in this study has a network depth of $N = 6$, yielding a total of $\lvert \mathcal{F} \rvert = 382$ parameters. The end-to-end downstream learning process can be expressed as

\begin{equation}\label{eq:end_to_end_mapping}
(\mathbf{Q})
\xmapsto[\mathcal{M}_{\text{FM}}(\mathbf{Q})]{}
(\mathcal{F}, \mathbf{C}^{\text{crystal}})
\xmapsto[\mathcal{H}_{\text{ODMN}}(\mathcal{F}, \mathbf{C}^{\text{crystal}})]{}
\bar{\mathbf{C}},
\end{equation}
where $\mathbf{Q}$ denotes the quaternion field of the voxel-based microstructure with dimensions $45 \times 45 \times 45 \times 4$.
The mapping $\mathcal{M}_{\text{FM}}$ represents the pretrained encoder equipped with a linear regression head, whose trainable parameters are optimized during fine-tuning to predict the ODMN parameter set $\mathcal{F}$. 
Given the predicted $\mathcal{F}$ and the single-crystal stiffness tensor $\mathbf{C}^{\text{crystal}}$, the analytical homogenization operator $\mathcal{H}_{\text{ODMN}}$ is then employed to compute the homogenized stiffness tensor $\bar{\mathbf{C}}$. 
This homogenized stiffness is the model output used to define the training loss, which is backpropagated through $\mathcal{H}_{\text{ODMN}}$ to update the parameters of $\mathcal{M}_{\text{FM}}$ during downstream fine-tuning.

The downstream training objective is defined as the relative Frobenius norm error between the ODMN-predicted homogenized stiffness and the corresponding ground truth obtained from direct numerical simulation (DNS),
\begin{equation}
    \mathcal{L}_{\text{ODMN}} 
    = \frac{1}{N_{\text{dataset}}} 
    \sum_{s=1}^{N_{\text{dataset}}} 
    \frac{
    \left\lVert 
    \bar{\mathbf{C}}^{\text{DNS}}_s 
    - 
    \bar{\mathbf{C}}_s
    \!\left(
    \mathbf{C}^{\text{crystal}}_s,
    \mathcal{F}_{s}
    \right)
    \right\rVert_{\mathrm{F}}
    }{
    \left\lVert 
    \bar{\mathbf{C}}^{\text{DNS}}_s 
    \right\rVert_{\mathrm{F}}
    },
    \label{eq:imn_loss}
\end{equation}
where $N_{\text{dataset}}$ denotes the total number of training samples, the subscript $s$ indexes the samples, $\mathcal{F}_{s}$ is the ODMN parameter set predicted for sample $s$, and $\lVert \cdot \rVert_{\mathrm{F}}$ is the Frobenius norm.

The choice of a linear projection from the latent descriptor to the ODMN parameters is deliberate and is not an unconstrained curve fit. The head is trained end-to-end \emph{through} the analytical, differentiable homogenization operator $\mathcal{H}_{\text{ODMN}}$ against direct-numerical-simulation stiffness, so the physics of hierarchical homogenization is embedded directly in the training signal (Eq.~\eqref{eq:imn_loss}); the linear map needs only to place the already-structured pretrained latent features into the ODMN parameter space. Keeping the head linear also minimizes the number of task-specific parameters, which is important in the label-scarce downstream regime.

During the online prediction stage, previously unseen microstructures are first processed by the trained mapping $\mathcal{M}_{\text{FM}}$ to predict the corresponding ODMN parameter set $\mathcal{F}$. The predicted $\mathcal{F}$ defines a standalone ODMN, which is subsequently combined with the constituent phase material behavior, such as crystal plasticity, to evaluate the nonlinear mechanical response, as illustrated in Fig.~\ref{fig:odmn_downstream}(b). This strategy enables efficient prediction of nonlinear mechanical behavior in complex polycrystalline microstructures, while preserving the physical consistency of the underlying multiscale framework.

\subsubsection{Dataset construction for downstream task II: Nonlinear response modeling}\label{sec:task2_dataset}

For the second downstream task, a labeled dataset was constructed to enable supervised training for ODMN parameter prediction. Following the strategy proposed by Dai et al.~\cite{dai2021studying}, a total of 1{,}600 polycrystalline microstructures were generated, comprising four representative crystallographic texture types with 400 samples per texture category.

For each microstructure, 500 distinct sets of single-crystal elastic stiffness parameters were generated to enrich the dataset and to cover a wide range of mechanical responses.
The dataset was randomly partitioned into training and validation subsets, with 320 and 80 microstructures per texture type assigned to the training and validation sets, respectively.

For each crystallographic texture instance, DREAM3D-NX was used to generate RVEs comprising approximately 810 equiaxed grains, discretized on a $45 \times 45 \times 45$ voxel grid. The four representative texture types considered in this study are summarized as follows:

\begin{itemize}\sloppy
    \item[(i)] {Strong-textured-1 (S1):} One dominant orientation with \emph{Weight} = 500,000 and \emph{Sigma} = 1; all other orientations are assigned \emph{Weight} = 1 and \emph{Sigma} = 1.
    \item[(ii)] {Strong-textured-2 (S2):} One dominant orientation with \emph{Weight} = 500,000 and \emph{Sigma} = 8; all other orientations are assigned \emph{Weight} = 1 and \emph{Sigma} = 1.
    \item[(iii)] {Weak-textured-1 (W1):} All orientations are uniformly assigned \emph{Weight} = 1 and \emph{Sigma} = 1, corresponding to a nearly random texture.
    \item[(iv)] {Weak-textured-2 (W2):} Two dominant orientations, each with \emph{Weight} = 500,000 and \emph{Sigma} = 10; all other orientations are assigned \emph{Weight} = 1 and \emph{Sigma} = 1.
\end{itemize}

\begin{figure}[ht]
\centering
\includegraphics[width=0.5\linewidth]{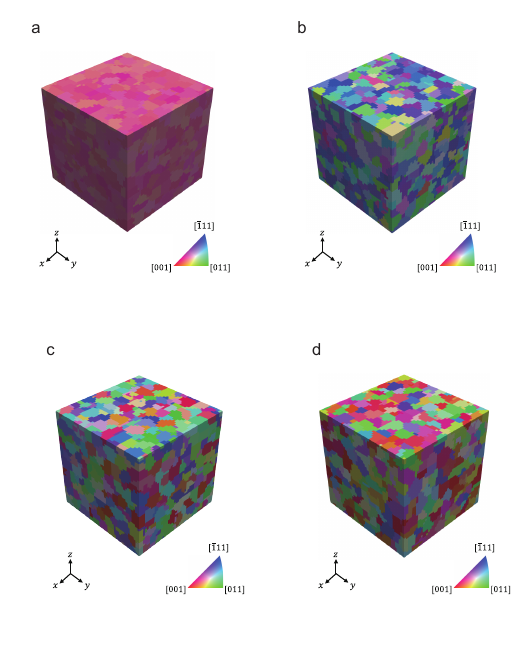}
\caption{Representative RVEs for the four texture types used in Downstream Task~II: (a) S1, (b) S2, (c) W1, and (d) W2, shown as IPF-colored orientation maps.}
\label{fig:representative_RVEs}
\end{figure}

The corresponding representative RVEs are shown in Fig.~\ref{fig:representative_RVEs}. The visual contrast among the four inverse pole figure (IPF) colored orientation maps shows the intended variation in texture strength and orientation dispersion, from nearly single-crystal-like configurations to nearly random polycrystals.

The four texture types considered here represent a subset, rather than an exhaustive coverage, of the texture hull. They are intentionally selected to capture polycrystalline microstructures with distinct texture intensities and orientation dispersions, providing a controlled yet representative testbed for downstream tasks.

After RVE realization, each microstructure is paired with 500 distinct single-crystal stiffness triplets
$\{ C^{\text{crystal}}_{11}, C^{\text{crystal}}_{12}, C^{\text{crystal}}_{44} \}$,
which together define the cubic single-crystal stiffness tensor $\mathbf{C}^{\text{crystal}}$, written in Voigt notation, used in Eq.~\eqref{eq:end_to_end_mapping}:

\begin{equation}\label{eq:crystal_stiffness_matrix}
\mathbf{C}^{\text{crystal}} =
\begin{bmatrix}
{C}^{\text{crystal}}_{11} & {C}^{\text{crystal}}_{12} & {C}^{\text{crystal}}_{12} & 0 & 0 & 0 \\
{C}^{\text{crystal}}_{12} & {C}^{\text{crystal}}_{11} & {C}^{\text{crystal}}_{12} & 0 & 0 & 0 \\
{C}^{\text{crystal}}_{12} & {C}^{\text{crystal}}_{12} & {C}^{\text{crystal}}_{11} & 0 & 0 & 0 \\
0 & 0 & 0 & {C}^{\text{crystal}}_{44} & 0 & 0 \\
0 & 0 & 0 & 0 & {C}^{\text{crystal}}_{44} & 0 \\
0 & 0 & 0 & 0 & 0 & {C}^{\text{crystal}}_{44}
\end{bmatrix}.
\end{equation}

The sampling strategy for the stiffness triplets
$\{ C^{\text{crystal}}_{11}, C^{\text{crystal}}_{12}, C^{\text{crystal}}_{44} \}$
follows the procedure detailed in the original ODMN study~\cite{Wei01},
ensuring sufficient variability in the elastic anisotropy of the constituent
single-crystal phases. Specifically, the three independent elastic constants
of the cubic single-crystal stiffness tensor were sampled as
\begin{equation}
\begin{aligned}
&C^{\text{crystal}}_{11},\,
C^{\text{crystal}}_{12},\,
C^{\text{crystal}}_{44}
\sim U\!\left(10^{-3},10^{3}\right)\ \text{GPa}, \\
&\text{subject to } C^{\text{crystal}}_{11}
-
C^{\text{crystal}}_{12}
>0 .
\end{aligned}
\label{eq:crystal_stiffness_sampling}
\end{equation}
Here, $U(a,b)$ denotes a uniform distribution over the interval $[a,b]$.
The constraint $C^{\text{crystal}}_{11}-C^{\text{crystal}}_{12}>0$,
together with the positive sampling range of
$C^{\text{crystal}}_{44}$, ensures the positive definiteness of the cubic
elastic stiffness tensor. For each pair of microstructure and sampled
single-crystal stiffness tensor, the corresponding homogenized stiffness
tensor $\bar{\mathbf{C}}^{\text{DNS}}$ was computed using the DAMASK-FFT
solver~\cite{roters2019damask}.

\section{Results and Discussion}\label{sec3}

\subsection{Pretraining performance and latent representation analysis}

During the pretraining stage, the foundation model is optimized in a self-supervised manner by randomly masking a subset of the voxel-based microstructure and reconstructing the missing regions.
The masking ratio thus constitutes a critical hyperparameter, as it directly controls both the difficulty of the reconstruction task and the amount of microstructural information available to the encoder.

\begin{figure}[ht]
    \centering
    \includegraphics[width=\linewidth]{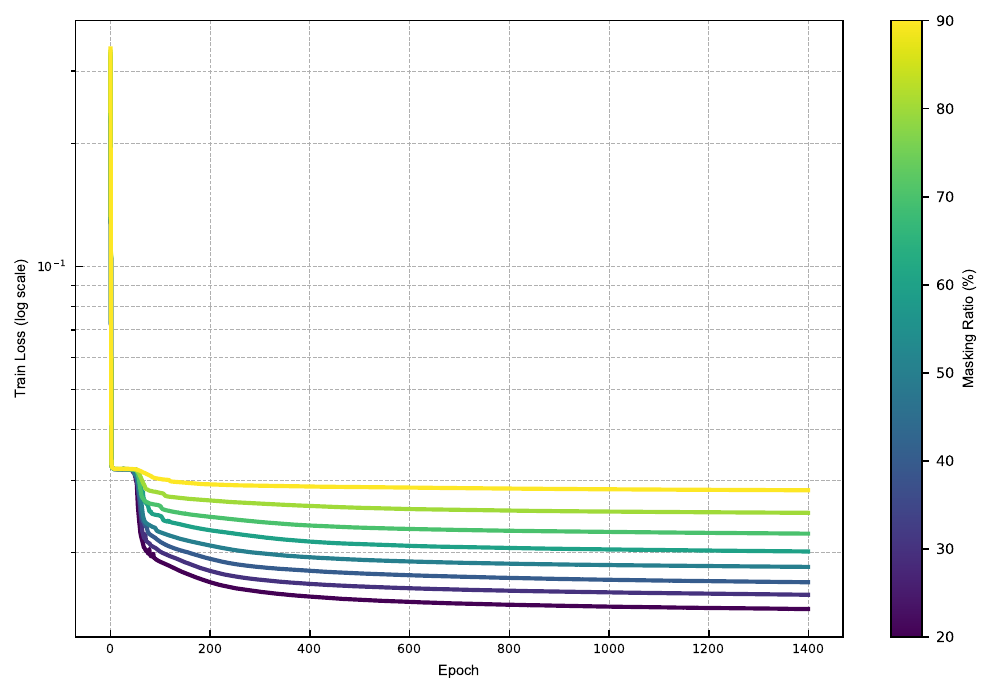}
    \caption{Pretraining loss curves for different masking ratios ranging from 20\% to 90\%. }
    \label{fig:pt_training_curve}
\end{figure}

To systematically assess its influence, an ablation study was conducted in which the masking ratio was varied from 20\% to 90\%, while all configurations were trained for 1400 epochs.
As shown in Fig.~\ref{fig:pt_training_curve}, all training curves exhibit monotonic decay and stable convergence, indicating well-behaved optimization dynamics across the entire range of masking ratios.
The final reconstruction error increases consistently with increasing masking ratio.
This trend is expected, since higher masking ratios reduce the number of visible patches and consequently limit the microstructural information available for exploitation during reconstruction.
Accordingly, the lower reconstruction errors observed at small masking ratios should be attributed to increased information availability rather than interpreted as evidence of a more expressive latent representation.

These observations indicate that reconstruction loss alone is insufficient for assessing representation quality.
A more informative evaluation requires examining how effectively the learned latent space transfers to downstream tasks, in which the ability to encode texture-sensitive and physically meaningful microstructural features is critical.
Downstream performance, therefore, provides a more practically relevant measure of the transferability of the pretrained encoder than reconstruction accuracy achieved during pretraining.

\newpage

The pretraining dataset was generated by sampling the texture hull with the HSS algorithm, which provides broad coverage of the high-dimensional crystallographic texture space.
Under this sampling strategy, a pretrained encoder that is both texture-aware and unbiased with respect to the data distribution is expected to produce a latent representation whose global distribution exhibits no artificial clustering or preferential concentration arising from the pretraining procedure itself.

\begin{figure}[ht]
    \centering
    \includegraphics[width=\linewidth]{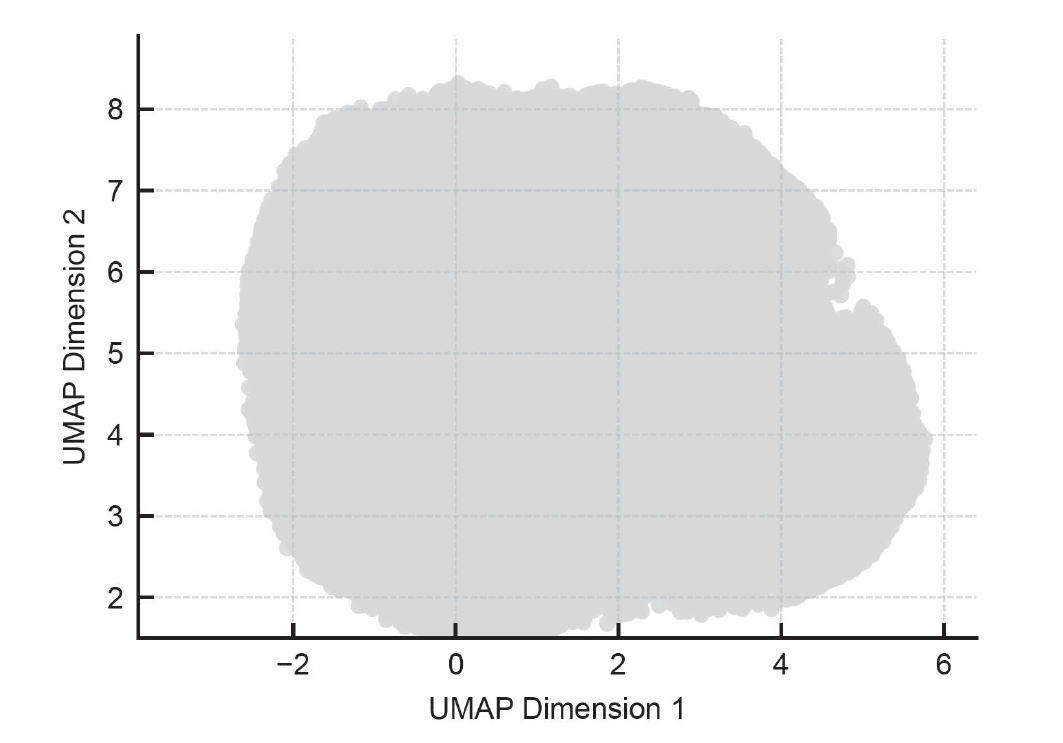}
    \caption{UMAP projection of the latent representations extracted from the pretrained encoder (masking ratio 40\%). Each gray point corresponds to a single RVE from the pretraining dataset. }
    \label{fig:pt_umap}
\end{figure}

To examine this behavior, the [CLS] token extracted from the encoder pretrained with a 40\% masking ratio for 1400 epochs was used as the latent descriptor for each RVE in the pretraining dataset.
These latent vectors were then projected onto a two-dimensional manifold using Uniform Manifold Approximation and Projection (UMAP) for visualization~\cite{mcinnes2018umap}.
As shown in Fig.~\ref{fig:pt_umap}, the resulting latent distribution appears broadly and smoothly distributed, without evident clustering or collapse into localized regions.
This observation indicates that the pretrained encoder maps the diverse set of crystallographic textures into a well-spread latent space, preserving continuity across the dataset.

Such a smoothly populated latent space suggests that the encoder avoids degenerative representations and maintains sufficient expressiveness to accommodate texture variability.
This property is particularly important for downstream transfer, where interpolation within the latent space is expected to correspond to meaningful variations in microstructural and texture-related features.

\newpage

\subsection{Downstream task I: stiffness prediction performance}

This downstream task evaluates the transferability of the pretrained encoder by predicting the three principal components of the homogenized stiffness tensor, $\bar{C}_{1111}$, $\bar{C}_{2222}$, and $\bar{C}_{3333}$.  
To quantify the effect of self-supervised pretraining, models initialized from encoders pretrained with different masking ratios were compared against a baseline model trained from scratch, i.e., without pretraining.

\begin{figure}[!htbp]
    \centering
    \includegraphics[width=\linewidth]{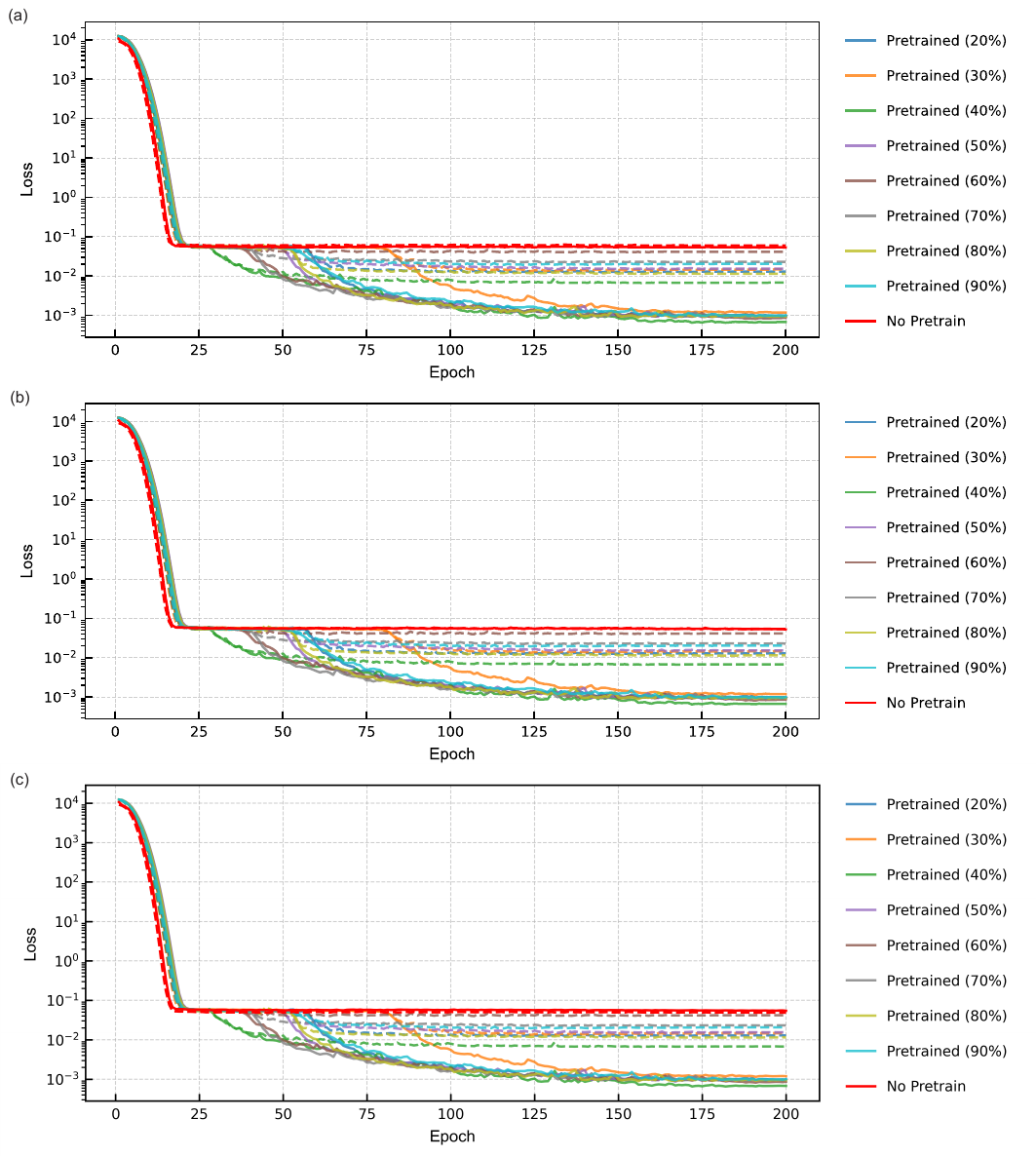}
    \caption{Training and validation loss curves for downstream stiffness prediction of (a) $\bar{C}_{1111}$, (b) $\bar{C}_{2222}$, and (c) $\bar{C}_{3333}$ under different masking ratios. Solid lines indicate training loss, whereas dashed lines indicate validation loss. }
    \label{fig:stiffness_training_curves}
\end{figure}

Figure~\ref{fig:stiffness_training_curves} shows the evolution of the training and validation losses for the three stiffness components.
Across all masking ratios, the fine-tuning process exhibits stable optimization behavior, with both losses decreasing consistently throughout training.
A clear and systematic performance gap emerges when comparing pretrained models with the non-pretrained baseline.
For all stiffness components and across all masking configurations, the pretrained models achieve substantially lower prediction errors than the model trained without pretraining.
This result indicates that the latent representation learned during self-supervised pretraining provides an effective inductive structure for predicting texture-dependent elastic properties.

\newpage

\begin{figure}[ht]
    \centering
    \includegraphics[width=\linewidth]{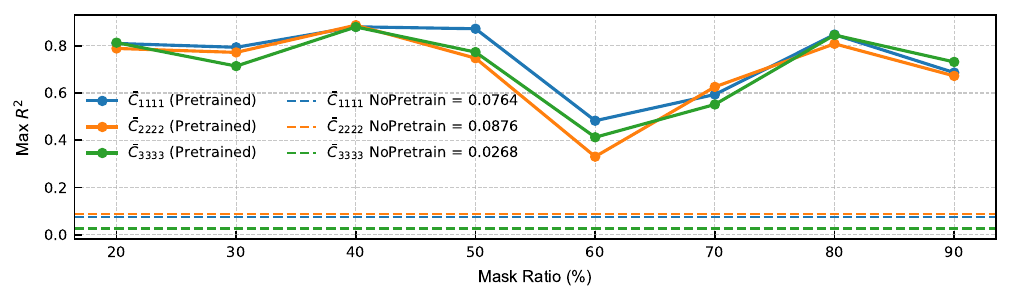}
    \caption{Validation $R^2$ scores for predicting the stiffness components $\bar{C}_{1111}$, $\bar{C}_{2222}$, and $\bar{C}_{3333}$ using models trained with and without self-supervised pretraining. Results are shown for pretrained encoders with different masking ratios.}
    \label{fig:FT_multi_Cs}
\end{figure}

A quantitative comparison of predictive accuracy is presented in Fig.~\ref{fig:FT_multi_Cs}, which reports the validation $R^2$ values for the three stiffness components.
The baseline model trained from scratch achieves $R^2$ values below 0.09 across all components.
Given that the dataset spans the entire texture hull, resulting in pronounced anisotropy and large stiffness variations, such limited performance is expected in the absence of informative prior structural representations.

In contrast, all pretrained configurations achieve substantially higher predictive accuracy, even under the data-limited conditions of this downstream task.
Among them, the model pretrained with a masking ratio of 40\% attains the best overall performance, with validation $R^2$ values exceeding 0.8 for all three stiffness components.
These results show that self-supervised pretraining provides informative latent representations.

In summary, three key observations can be drawn from these results.
First, self-supervised pretraining yields latent representations that transfer effectively to elastic property prediction.
Second, the masking ratio influences the expressiveness and utility of the learned embedding, leading to measurable differences in downstream accuracy.
Third, pretraining substantially improves regression performance relative to models trained without it, which matters for data-efficient microstructure--property learning.

\newpage

\subsection{Downstream task II: nonlinear response prediction}\label{sec:task2_results}

This downstream task evaluates the transferability of the pretrained encoder by assessing its ability to infer ODMN parameters during offline training and predict nonlinear mechanical responses during online prediction.
Accordingly, the evaluation comprises two complementary aspects.
The first focuses on the fine-tuning behavior of the pretrained encoder when coupled with the ODMN.
The second examines the accuracy of the inferred ODMNs in predicting nonlinear stress--strain responses for previously unseen microstructures.

During the offline stage, the pretrained encoder was integrated with the ODMN and optimized end-to-end to infer ODMN parameters that approximate the homogenized stiffness responses used for supervised training.
Pretrained encoders with masking ratios ranging from 20\% to 90\% were considered, together with an additional baseline model trained entirely from scratch.

\begin{figure}[ht]
    \centering
    \includegraphics[width=\linewidth]{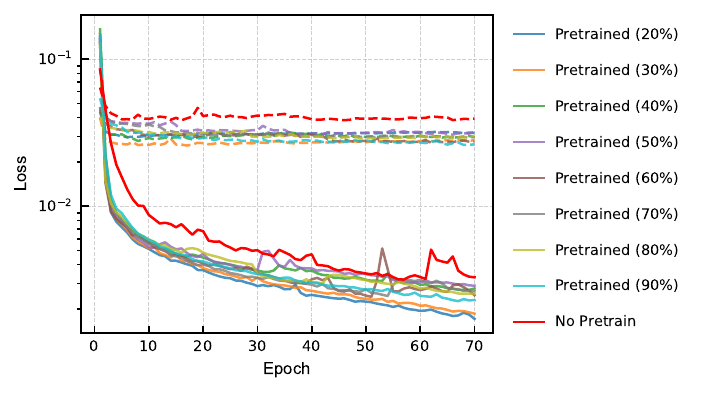}
    \caption{Training and validation loss curves for the downstream task of predicting ODMN parameters under different masking ratios during offline end-to-end fine-tuning. Solid lines denote training loss, whereas dashed lines denote validation loss.}
    \label{fig:odmn_training_curves}
\end{figure}

As shown in Fig.~\ref{fig:odmn_training_curves}, all configurations exhibit smooth and stable convergence. However, the model trained from scratch yields the highest training error and the largest validation error among all cases, and exhibits the widest train--validation gap. This behavior suggests that, in the absence of pretrained structural priors, the encoder is initialized farther from a microstructural representation space that is effective for ODMN parameter inference. Consequently, the supervised optimization converges to a less favorable solution, resulting in both poorer fitting of the training data and weaker generalization to validation microstructures. In contrast, masked pretraining provides an encoder representation that is already informed by the underlying microstructural features, thereby reducing both training and validation errors while narrowing the generalization gap.

\begin{figure}[ht]
    \centering
    \includegraphics[width=\linewidth]{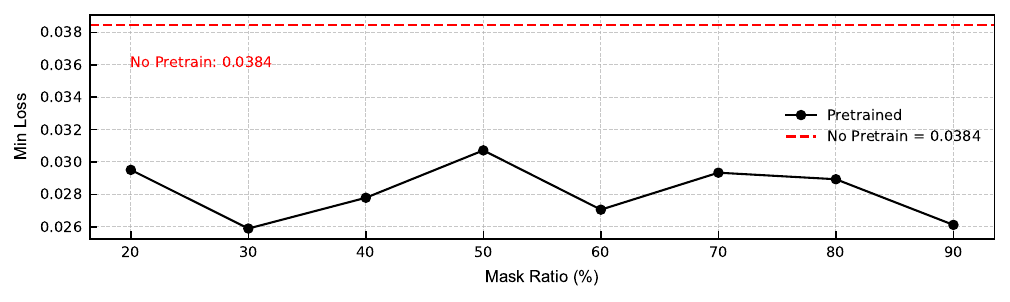}
    \caption{Minimum validation loss for predicting ODMN parameters under different masking ratios during offline end-to-end fine-tuning.}
    \label{fig:odmn_training_Summary}
\end{figure}

The validation results further confirm the importance of pretraining. As summarized in Fig.~\ref{fig:odmn_training_Summary}, the encoder pretrained with a masking ratio of 30\% achieves the lowest validation error of 2.59\%, whereas the model trained from scratch yields a validation error of 3.84\%. This corresponds to a relative reduction of approximately 33\%, demonstrating that masked pretraining substantially improves the generalization capability of the ODMN parameter inference model.

During the online prediction stage, the pretrained model with a 30\% masking ratio was selected for evaluation, as it demonstrated the best performance in the offline ODMN training stage.
Four representative RVEs were constructed for testing, corresponding to the texture types S1, S2, W1, and W2 defined in Section~\ref{sec:task2_dataset}.

For each microstructure, the trained encoder was used to infer the parameters of a standalone ODMN, which was then coupled with a crystal-plasticity model to predict the uniaxial loading--unloading--reloading stress--strain response~\cite{roters2019damask, Wei01}. Details of the crystal-plasticity formulation and elastic response are provided in Appendices~\ref{appendixA} and~\ref{appendixB}, respectively, and the material parameters used in the simulations are summarized in Table~\ref{tab:crystal_plasticity_parameters}. Reference solutions were obtained from full-field DNS using the DAMASK-FFT solver~\cite{roters2019damask}.

\begin{figure}[!ht]
    \centering
    \includegraphics[width=\linewidth]{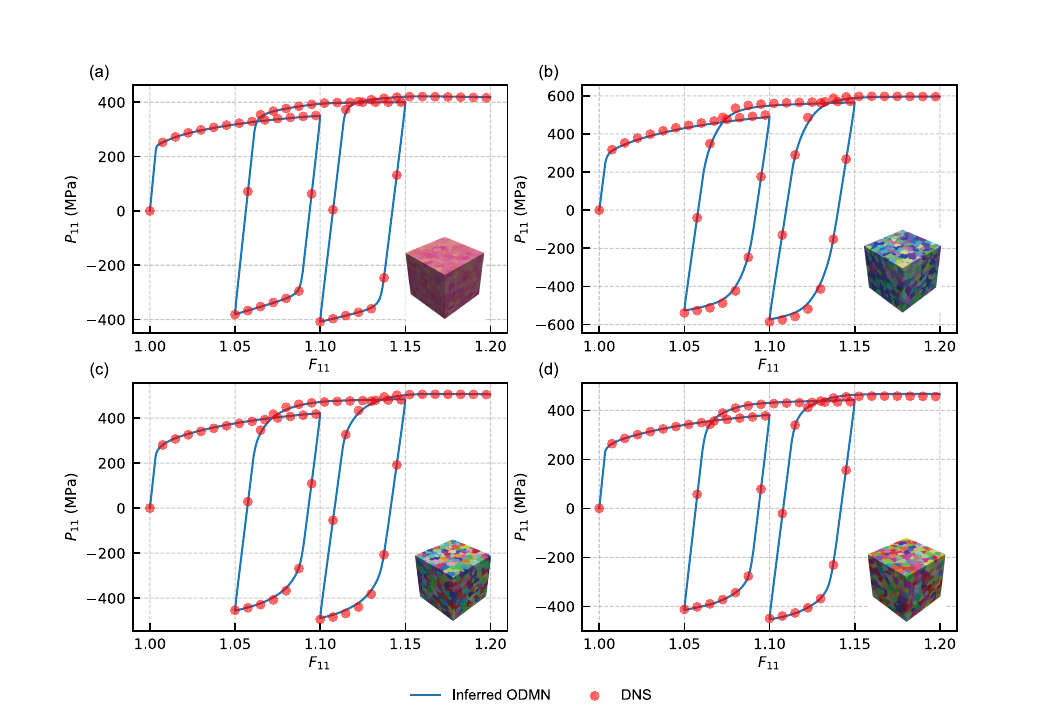}
    \caption{Predicted stress--strain responses obtained from standalone ODMNs inferred for four previously unseen RVEs subjected to cyclic loading. The four cases correspond to (a) S1, (b) S2, (c) W1, and (d) W2.}
    \label{fig:ss_curve_cyclic}
\end{figure}

The resulting stress--strain responses are presented in Fig.~\ref{fig:ss_curve_cyclic}.
For all four microstructures, the ODMN-predicted curves show close agreement with the DNS reference results along the entire loading--unloading--reloading path.
To quantitatively assess the prediction accuracy, two normalized error metrics were adopted following established definitions in the literature~\cite{huang2022microstructure}.
The mean relative error characterizes the average deviation between the predicted and DNS stresses over the loading history, while the maximum relative error captures the largest discrepancy observed along the entire loading path.
These metrics are defined as

\begin{equation}
\text{mean-relative error} =
\frac{ \frac{1}{n} \sum_{i=1}^{n} \left| P_i^{\text{DNS}} - P_i^{\text{ODMN}} \right| }
     { \max_{i=1,\ldots,n} \left| P_i^{\text{DNS}} \right| } ,
\label{eq:mean_rel_error}
\end{equation}

\begin{equation}
\text{max-relative error} =
\frac{ \max_{i=1,\ldots,n} \left| P_i^{\text{DNS}} - P_i^{\text{ODMN}} \right| }
     { \max_{i=1,\ldots,n} \left| P_i^{\text{DNS}} \right| } ,
\label{eq:max_rel_error}
\end{equation}

\noindent where $P_i^{\text{DNS}}$ and $P_i^{\text{ODMN}}$ denote the homogenized stress components obtained from DNS and from the inferred ODMN at the $i$-th point of the loading history, and $n$ is the total number of points along the path.

\begin{table}[htbp]
\centering
\renewcommand{\arraystretch}{1.3}
\caption{Relative stress prediction errors of inferred ODMNs under cyclic loading, compared against DNS results.}
\begin{tabular}{lcccc}
    \hline
    RVE & S1 & S2 & W1 & W2 \\
    \hline
    mean-relative error (\%) & 0.37 & 1.97 & 1.02 & 0.72  \\
    max-relative error (\%)  & 2.87 & 5.63 & 3.60 & 2.25  \\
    \hline
\end{tabular}
\label{tab:ss_curve_error_metrics_cyclic}
\end{table}

The quantitative results are summarized in Table~\ref{tab:ss_curve_error_metrics_cyclic}.
Across all four RVEs, the mean relative error remains below 2\%, indicating that the inferred ODMNs reproduce the nonlinear stress response with consistently high accuracy.
The largest maximum relative error is observed for RVE~S2, reaching 5.63\%.
This deviation corresponds to a slight discrepancy relative to the DNS reference and is limited in magnitude compared to the overall nonlinear response.

These results show that coupling the foundation model with the ODMN enables accurate prediction of nonlinear mechanical responses for previously unseen microstructures spanning a wide range of texture characteristics.

\begin{table}[htbp]
\centering
\caption{Elastic and plastic material parameters for AA6022-T4 \cite{barrett2019deep, damask_documentation}.}
\label{tab:crystal_plasticity_parameters}
\renewcommand{\arraystretch}{1.2}
\setlength{\tabcolsep}{6pt}
\begin{tabular}{cccccccc}
\hline
\textbf{$N_s$} & \textbf{$h_0^{\text{sl-sl}}$}(GPa) & \textbf{$\xi^{\alpha}_{\infty}$}(MPa) & \textbf{$\xi^{\alpha}_{0}$}(MPa) & \textbf{$n$} & \textbf{$a$} & \textbf{$\dot{\gamma}_0$}$(\text{s}^{-1})$ & \textbf{$h_{\text{int}}^{\alpha}$} \\
\hline
12 & 1.02 & 266 & 76 & 20 & 3.7 & 0.001 & 0 \\
\hline
\end{tabular}

\vspace{0.5cm}

\begin{tabular}{cccc}
\hline
{$C_{11}$ (GPa)} & {$C_{12}$ (GPa)} & {$C_{44}$ (GPa)} & {$h^{\text{sl-sl}}$} \\
\hline
107.3 & 60.8 & 28.3 & [1, 1, 5.123, 0.574, 1.123, 1.123, 1] \\
\hline
\end{tabular}
\end{table}

\newpage

\newpage

\subsection{Computational cost and efficiency}\label{sec:cost}

A central practical question is the offline computational cost of training the proposed pipeline and the online efficiency gained once the inferred ODMN is available. We therefore report the measured one-time training costs separately from the recurring online prediction cost incurred for each previously unseen microstructure. The offline costs reported here include only the GPU time used for self-supervised pretraining and downstream fine-tuning; they do not include the cost of microstructure generation or DNS-based label generation. Accordingly, the following comparison should be interpreted as a training-and-inference cost summary, rather than a full end-to-end accounting of dataset construction.

The measured costs are summarized in Table~\ref{tab:cost}. Pretraining a single encoder on $100{,}000$ RVEs required approximately $45.6$~GPU-h for the 30\%-masked encoder (the best-performing encoder in Downstream Task~II) and approximately $45.8$~GPU-h for the 40\%-masked encoder (the best-performing encoder in Downstream Task~I); the pretraining cost is nearly insensitive to the masking ratio. For Downstream Task~I, the reported fine-tuning cost of approximately $1.9$~GPU-h denotes the average training time for one stiffness-component predictor. Thus, training the three predictors for $\bar{C}_{1111}$, $\bar{C}_{2222}$, and $\bar{C}_{3333}$ requires approximately $3\times1.9\approx5.7$~GPU-h in total. For Downstream Task~II, fine-tuning the encoder and task-specific head required approximately $38.6$~GPU-h, using $1{,}600$ RVEs with $500$ sampled stiffness-matrix instances per RVE.

For online prediction in Downstream Task~II, the inferred standalone ODMN requires $2{,}822$~CPU-s per unseen microstructure on average, whereas the corresponding full-field crystal-plasticity DNS using DAMASK-FFT requires $859{,}068$~CPU-s under the same nonlinear loading history. This gives a like-for-like online CPU-time speedup of
\[
\frac{859{,}068}{2{,}822} \approx 304.4 .
\]
This comparison isolates the recurring online evaluation cost and shows that, once trained, the inferred ODMN substantially reduces the cost of predicting nonlinear responses for new microstructures.

\begin{table}[htbp]
\centering
\begin{threeparttable}
\caption{Computational cost breakdown for offline training and online prediction. GPU-hours and CPU-seconds are reported separately and should not be summed directly. Online evaluations were performed on Intel\textsuperscript{®} Xeon\textsuperscript{®} Platinum 8480 (2.0\,GHz) CPUs; the GPU hardware for each training stage is listed in Table~\ref{tab:training_config}.}
\label{tab:cost}
\footnotesize
\renewcommand{\arraystretch}{1.2}
\setlength{\tabcolsep}{3pt}
\begin{tabular}{@{}lll@{}}
\toprule
Stage & Quantity & Cost \\
\midrule
\multicolumn{3}{l}{\emph{One-time offline training}} \\
\quad Self-supervised pretraining
    & $100{,}000$ RVEs, 30\% masking
    & $\sim$45.6 GPU-h \\
\quad Self-supervised pretraining
    & $100{,}000$ RVEs, 40\% masking
    & $\sim$45.8 GPU-h \\
    
\quad Downstream Task~I 
    & $5{,}000$ RVEs 
    & $\sim$1.9 GPU-h\tnote{a} \\
    
\quad Downstream Task~II 
    & $1{,}600$ RVEs\tnote{b} 
    & $\sim$38.6 GPU-h \\
\midrule
\multicolumn{3}{l}{\emph{Per unseen microstructure: Task~II online prediction}} \\
\quad Inferred ODMN evaluation 
    & average over 4 RVEs 
    & $2{,}822$ CPU-s \\
    
\quad Crystal-plasticity DNS (DAMASK-FFT) 
    & average over 4 RVEs 
    & $859{,}068$ CPU-s \\
    
\quad \textbf{Speedup} 
    & -- 
    & $\mathbf{304.4\times}$ \\
\bottomrule
\end{tabular}
\begin{tablenotes}
\footnotesize
\item[a] The reported cost is the average fine-tuning time for one stiffness-component predictor. 
\item[b] Each RVE is associated with $500$ sampled stiffness-matrix instances, giving $800{,}000$ training instances in total.
\end{tablenotes}
\end{threeparttable}
\end{table}

\subsection{Model transferability, interpretability, and current limitations}\label{sec:limits}

The capabilities of the proposed foundation model are demonstrated through two downstream tasks characterized by distinct physical settings.
In Task~I, the objective is to predict the homogenized elastic stiffness components.
Although the target response is linear elastic, the input space spans the full crystallographic texture hull, leading to substantial microstructural variability despite the limited availability of labeled data.
In Task~II, the evaluation is conducted in two stages: an offline training stage and an online prediction stage.
During the offline stage, the pretrained encoder is integrated with the ODMN to infer the complete set of ODMN parameters.
The model is optimized using labeled homogenized stiffness data to learn the mapping from voxelized microstructures to the ODMN parameterization.
Although the training microstructures in Task~II cover only a subset of the texture hull, the task remains considerably more challenging because it requires learning a mapping from microstructures to the corresponding ODMN parameters.
During the online prediction stage, the fine-tuned encoder is applied to previously unseen microstructures to infer standalone ODMNs, which subsequently perform nonlinear extrapolation through crystal-plasticity-based homogenization.
Task~II thus demonstrates both the transferability of the pretrained encoder and the ability of the inferred ODMNs to generalize beyond the linear training regime and perform nonlinear mechanical extrapolation.

To further elucidate the role of pretraining, a non-pretrained encoder was included as a baseline in both downstream tasks.
Across all masking ratios, the pretrained encoders consistently outperform the non-pretrained baseline, with the largest gap observed during the offline training stage of Task~II (Section~\ref{sec:task2_results}).
Self-supervised pretraining organizes the latent space into a physically structured manifold that captures essential microstructural statistics.
This provides a useful inductive bias that steers optimization toward stable and transferable solutions.
Such behavior is particularly important in materials informatics, where labeled data are inherently scarce and generalization to previously unseen microstructures is essential.

Despite these encouraging results, the present study is a deliberate first step, and its scope should be read accordingly.
(i) Pretraining and evaluation use synthetic microstructures from DREAM3D-NX and labels from the DAMASK-FFT solver; real materials additionally exhibit grain-boundary character, dislocation substructure, second phases, and defects, so a synthetic-to-real domain gap remains, and experimental validation on FCC alloys, for example, through uniaxial tensile and cyclic loading tests, is the priority next step.
(ii) Grains are assumed equiaxed; processing routes such as rolling, extrusion, and additive manufacturing produce morphological texture (elongated, anisotropic grains) that the current corpus does not cover.
(iii) The corpus is restricted to FCC; body-centered cubic and hexagonal close-packed systems, with distinct slip families and anisotropy, are required to claim crystallographic generality.
(iv) Two downstream tasks are demonstrated; broader claims of universality would require additional tasks, such as conductivity estimation or phase and defect classification.
(v) A fixed $45\times45\times45$ resolution under periodic boundary conditions is used; the sensitivity of the learned representations and predictions to grid resolution and to non-periodic boundary conditions relevant to localized components remains to be characterized.
We regard these as well-defined directions rather than obstacles: the open dataset, code, and pretrained encoder are released to enable the community to extend the corpus along each axis.

\section{Conclusions}\label{sec4}

This study introduced a three-dimensional polycrystal foundation model pretrained on a large corpus of FCC microstructures whose crystallographic orientations span the texture hull.
Through a self-supervised masked reconstruction strategy with masking ratios ranging from 20\% to 90\%, the encoder was trained on $100{,}000$ voxel-based microstructures and learned a physically meaningful latent representation. The resulting latent distribution is broadly and smoothly populated, indicating a continuous embedding of the pretraining dataset.

The effectiveness of the foundation model was demonstrated on two downstream tasks with distinct physical settings.
In Task~I, the pretrained encoder was employed as a feature extractor for homogenized stiffness prediction, consistently achieving higher accuracy than the non-pretrained baseline across all masking ratios.
In Task~II, the encoder was integrated with the ODMN to infer complete sets of ODMN parameters for nonlinear constitutive modeling.
In this setting, pretraining consistently reduced the prediction error of the inferred ODMN parameters, while the inferred ODMNs enabled accurate prediction of nonlinear responses for previously unseen microstructures via crystal-plasticity-based homogenization.

Together, these results show that the proposed foundation model transfers well across tasks and learns a physically structured latent representation of polycrystalline microstructures.
Such characteristics are particularly advantageous in data-scarce scientific regimes, where labeled microstructural data are costly to acquire and physics-consistent generalization is critical.
The foundation model can also be integrated with experimentally derived datasets.
As a latent-space bridge, it allows limited experimental observations to inform microstructure--property inference and strengthens the link between physical measurements and model predictions.

\section*{Acknowledgements}
This work is supported by the National Science and Technology Council, Taiwan, under Grant 111-2221-E-002-054-MY3, 112-2221-E-007-028, and 114-2221-E-002-010-MY3. We are grateful for the computational resources and support from the National Center for Research on Earthquake Engineering (NCREE), NIAR, Taiwan, NTUCE-NCREE Joint Artificial Intelligence Research Center, and the National Center of High-performance Computing (NCHC). The authors also wish to thank José Niño and Oliver K. Johnson for their valuable assistance with implementing the hierarchical simplex sampling and texture hull framework.

\appendix
\section{Phenomenological Crystal Plasticity Model}
\label{appendixA}

The local constitutive response invoked by ODMN during online prediction is governed by a phenomenological crystal plasticity law, formulated consistently with the DAMASK implementation and excluding deformation-twinning mechanisms~\cite{roters2019damask}.

Plastic flow is characterized through the plastic velocity gradient $\mathbf{L}_p$, obtained by summing the shear contributions of every active slip system $\alpha$:
\begin{equation}
    \mathbf{L}_p = \sum_{\alpha} \dot{\gamma}^\alpha (\mathbf{s}^\alpha_s \otimes \mathbf{n}^\alpha_s)
\end{equation}
\noindent Here $\mathbf{s}^\alpha_s$ and $\mathbf{n}^\alpha_s$ denote, respectively, the slip direction and the normal to the slip plane, while $\dot{\gamma}^\alpha$ is the slip rate on system $\alpha$.

The hardening of the slip resistance $\xi^\alpha$ describes its transition from an initial value $\xi_0^\alpha$ toward a saturation state $\xi_{\infty}^{\alpha}$, governed by
\begin{equation}
    \dot{\xi}^\alpha = h_0^{\text{sl-sl}}(1+h_{\text{int}}^\alpha) \times\sum_{\alpha'}^{N_s}\left | \dot{\gamma}^{\alpha'} \right |{\left | 1-\frac{\xi ^{\alpha'}}{\xi ^{\alpha'}_\infty} \right |}^{a}\text{sgn}\left(1-\frac{\xi ^{\alpha'}}{\xi ^{\alpha'}_\infty}\right)h^{\alpha\alpha'}
\end{equation}
in which $N_s$ is the total number of slip systems considered, $h_0^{\text{sl-sl}}$ a reference hardening modulus, $h_{\text{int}}^{\alpha}$ an interaction-hardening coefficient, $a$ the hardening exponent, and $h^{\alpha\alpha'}$ the matrix describing latent-hardening interactions among systems.

A power-law flow rule relates the slip rate of each system to the ratio between the resolved shear stress $\tau^\alpha$ and the slip resistance $\xi^\alpha$:
\begin{equation}
    \dot{\gamma}^\alpha = \dot{\gamma}^\alpha_0 {\left | \frac{\tau^\alpha}{\xi^\alpha} \right |}^n \text{sgn}(\tau^\alpha)
\end{equation}
where $\dot{\gamma}^\alpha_0$ is a reference shear rate and $n$ the strain-rate sensitivity exponent.

The resolved shear stress on each system follows from Schmid's law applied to the Mandel stress $\mathbf{M}^p$:
\begin{equation}
    \tau^\alpha = \mathbf{M}^p \cdot (\mathbf{s}^\alpha_s \otimes \mathbf{n}^\alpha_s)
\end{equation}

\section{Generalized Hooke's Law}
\label{appendixB}
The elastic response within the crystal-plasticity formulation adopted by ODMN is captured by a generalized Hooke's law, which relates the second Piola--Kirchhoff stress tensor $\mathbf{S}$ to the Green--Lagrange strain tensor $\mathbf{E}$ through the fourth-order elastic stiffness tensor $\mathbb{C}$:
\begin{equation}
    \mathbf{S} = \mathbb{C} : \mathbf{E}.
\end{equation}
\noindent Here $\mathbb{C}$ encodes the material's elastic moduli, and the operator $:$ denotes the double contraction between two tensors.

The Green--Lagrange strain is in turn defined in terms of the elastic part of the deformation gradient $\mathbf{F}^e$:
\begin{equation}
    \mathbf{E} = \frac{1}{2} \left( (\mathbf{F}^e)^\top \mathbf{F}^e - \mathbb{I} \right),
\end{equation}
\noindent where $\mathbb{I}$ denotes the second-order identity tensor.

\section*{Declarations}
The authors declare that there are no competing interests.

\section*{Data Availability}
The datasets used in this study are publicly available at the Zenodo repository: \url{https://doi.org/10.5281/zenodo.18009973}.

\section*{Code Availability}
{\sloppy
The code used in this study can be accessed at the following GitHub repository: \url{https://github.com/BerryWei/Foundation-Model-for-Polycrystalline-Material-Informatics}.
\par}

\section*{Author Contributions}
Ting-Ju Wei: Writing – original draft, Visualization, Software, Methodology, Investigation, Formal analysis, Data curation, Conceptualization.
Chuin-Shan Chen: Writing – review \& editing, Supervision, Resources, Project administration, Funding acquisition, Conceptualization.

\bibliographystyle{unsrt}  
\bibliography{references}






\end{document}